\documentclass[a4paper,10pt,numbers=noenddot]{scrartcl}
\usepackage{graphicx}
\usepackage[ansinew]{inputenc}
\usepackage{immd10_thesis}
\usepackage{color}
\usepackage[USenglish]{babel}
\usepackage[loose]{subfigure}
\usepackage{amsmath}
\usepackage{url}
\usepackage{ifthen}
\usepackage{multirow}
\usepackage{rotating}
\usepackage{listings}
\usepackage[htt]{hyphenat}
\usepackage[small,bf]{caption}

\newcommand{\lineref}[1]{\hyperref[#1]{line~\ref*{#1}}}
\newcommand{\algref}[1]{\hyperref[#1]{Algorithm~\ref*{#1}}}

\DeclareMathAlphabet{\mathpzc}{OT1}{pzc}{m}{it}
\newcommand{\pe}{$\mathpzc{pe}$}

\newcommand{\blaze}{{\it Blaze}}

\DeclareMathAlphabet{\mathpzc}{OT1}{pzc}{m}{it}
\newcommand{\ublas}{{\it uBLAS}}

\DeclareMathAlphabet{\mathpzc}{OT1}{pzc}{m}{it}
\newcommand{\blitz}{{\it Blitz++}}

\newcommand{\ourtitle}{Expression Templates Revisited: A Performance Analysis of the Current ET Methodology}

\usepackage[pdftex,hyperindex,plainpages=false,pdfpagelabels]{hyperref}
\hypersetup{pdftitle={\ourtitle},
            pdfauthor={Klaus Iglberger},
            pdfsubject={Lehrstuhlbericht},
            colorlinks=true,
            linkcolor=blue,
            citecolor=red,
            bookmarksnumbered=true,
            bookmarksopenlevel=1,
            bookmarksopen=true,
            linktocpage=false,
            pdfpagelayout=OneColumn
            }

\graphicspath{{pics/}}

\definecolor{gray}{rgb}{0.95,0.95,0.95}
\definecolor{lightgray}{gray}{0.45}
\definecolor{myblue}{rgb}{0.22,0.22,0.61}
\definecolor{myred}{rgb}{0.6,0,0}

\begin{document}

\title{\ourtitle}

\author{\vspace{0.25cm}\\
Klaus Iglberger$^\star$, Georg Hager$^\dagger$, Jan Treibig$^\dagger$, Ulrich R{\"u}de$^\ast\;^\star$\\ \vspace{-0.1cm}\\
~\\
$^\star$ Central Institute for Scientific Computing\\
 Friedrich-Alexander University of Erlangen-Nuremberg\\
 91058 Erlangen, Germany\\
~\\
$^\dagger$ Erlangen Regional Computing Center\\
 Friedrich-Alexander University of Erlangen-Nuremberg\\
 91058 Erlangen, Germany\\
~\\
$^\ast$ Chair for System Simulation\\
 Friedrich-Alexander University of Erlangen-Nuremberg\\
 91058 Erlangen, Germany
}


\maketitle
\thispagestyle{empty}
%
%
%
%
%
%
%
%
%
%
%
%


{\bf Abstract:}\\
\begin{abstract}

In the last decade, Expression Templates (ET) have gained
a reputation as an efficient performance optimization tool for C++
codes. This reputation builds on several ET-based linear algebra
frameworks focused on combining both elegant and high-performance C++
code. However, on closer examination the assumption that ETs are
a performance optimization technique cannot be maintained. In this
paper we demonstrate and explain the inability of current ET-based frameworks
to deliver high performance for dense and sparse linear algebra
operations, and introduce a new ``smart'' ET implementation that truly
allows the combination of high performance code with the elegance and
maintainability of a domain-specific language.

\end{abstract}
~\\
{\bf Keywords:}\\
\newline
Expression Templates, Performance Optimization, High Performance Programming,
Linear Algebra, Boost, \ublas, \blitz, \blaze



\section{Introduction}
\label{sec:introduction}

Expression Templates (ETs) as originally introduced by
Veldhuizen in 1995~\cite{veldhuizen95,veldhuizen96} are intended to be
a ``performance optimization for array-based operations.'' The general
goal is to avoid the unnecessary creation of temporary objects during
the evaluation of arithmetic expressions with overloaded operators in
C++. Commonly demonstrated using simple ${\cal O}(n)$ array operations
like additions, it achieves performance levels similar to hand-crafted
C code while maintaining an elegant mathematical syntax. This success
led to quick adoption in standard
textbooks~\cite{vandevoorde:2003:C++,Abrahams:2005:C++TMP}, and ETs
are thus widely accepted as \emph{the} technique for high-performance
array math in C++.

Widely known libraries that fully implement ET-based arithmetics are
\blitz~\cite{blitz}, which was developed as ``a C++ class library for
scientific computing which provides performance on par with Fortran
77/90,'' and Boost \ublas~\cite{boost_ublas}, which is part of the
Boost project~\cite{boost}. Both frameworks successfully use ET concepts
to avoid the creation of temporaries. They provide fast array arithmetic
and still (mostly) maintain an intuitive, mathematical syntax via C++
operators. Also, both frameworks extend the ET methodology to matrices
and provide BLAS level 2 and 3 operations. In comparison to \blitz,
Boost \ublas{} extends the idea of ETs to sparse vectors and matrices.

The starting point of this paper is an evaluation of the single-core
(serial) performance of the \blitz{} and Boost \ublas{} libraries in the
context of high performance computing (HPC). Although the idea of
expression templates has a wider scope (they are, e.g., used for lambda
expressions~\cite{boost_lambda}), we focus on their performance aspect
in the context of numerical libraries. Based on those results we will
explain in detail why the current ET methodology is not suited for
high performance computing in general. As a solution we describe an
alternative ET approach, which combines the advantages of a high-level
language with architecture-specific performance optimization, and is
thus intrinsically suited for HPC. This ``smart'' ET methodology is
implemented in the \blaze{} library that was developed in context of
the \pe{} physics engine~\cite{iglberger:phd:10}. Note that we ignore
GPGPU computing altogether and focus on a contemporary CPU architecture,
the Intel ``Westmere.'' In order to demonstrate the achievable performance,
we will compare all results from the ET libraries to optimized BLAS code
(using the Intel MKL~\cite{mkl}).

The paper is organized as follows. In Section~\ref{sec:related_work}
we will give a short overview of related work, before
Section~\ref{sec:benchmark_platform} briefly summarizes the details
of our benchmark platform. Section~\ref{sec:idea} recapitulates the
current ET techniques and evaluates ET performance for the standard
benchmark (dense vector addition).
In Section~\ref{sec:mmm} we extend the analysis to dense
matrix-matrix multiplication and uncover some of the limitations of
standard ETs. We turn to study the use of ETs for sparse
data structures and complex expressions (operator chaining) in
Sections~\ref{sec:sparse} and \ref{sec:complex}. Finally we propose the
new methodology of ``Smart Expression Templates,'' which corrects the
problems of standard ETs by combining the
positive aspects of a domain-specific language with BLAS performance,
in Section~\ref{sec:smart_et}. Section~\ref{sec:inlining}
elaborates on the aspect of inlining in the context of ETs, before
Section~\ref{sec:conclusion} concludes the paper and provides
suggestions for future work.

\section{Related Work}
\label{sec:related_work}

Not many groups have invested work to look into the performance of ETs.
Bassetti~\cite{bassetti98} analyzed the performance of C++ expression templates in comparison
to Fortran 77 code. They show that the promise of performance of ETs is not uniformly guaranteed
across the different implementations of ETs, which they blame on the high demand on registers in
complex ET implementations. H\"ardtlein~\cite{haerdtlein09} introduced the concept of ``easy
expression templates'', which are easier to implement than classical ET, and the concept of
``fast expression templates'', which use static memory to improve the performance of array
operations.


\section{Benchmark Platform}
\label{sec:benchmark_platform}

A 6-core Intel Westmere CPU at 2.93\,GHz with 12\,MByte of shared L3 cache was used for
all benchmarks. The GNU g++ 4.4.2 and Intel 11.1 compilers produced very similar
performance results, so we always only present the results of the GNU g++ compiler.
To allow a direct comparison of the different ET methodologies, we do not
employ any low-level optimization apart from proper loop ordering, where appropriate.
\blitz, Boost \ublas, and \blaze{} were benchmarked as given. All results are normalized
to the fastest measured performance across the different frameworks for each particular
test case. For all test cases with dense vectors and matrices we additionally provide
MFlops/s values.



\section{The Idea Behind Expression Templates}
\label{sec:idea}


In this section we will recap the basic mechanisms at work in ETs. As an example, we will
use the addition of two dense vectors of type \lstinline|Vector|\footnote{We will focus on the
essential aspect of expression templates here and therefore omit all unnecessary details. For
instance, we are aware that the \lstinline|Vector| class could be implemented as a class template,
but this would unnecessarily bloat the code and obscure the core of ETs.}:

\begin{lstlisting}[numbers=left,
                   frame=tb,
                   caption={[Addition of two dense vectors.]
                             Addition of two dense vectors},
                   label={lst:idea_1}]
Vector a, b, c;
// ... Initialization of vector a and b
c = a + b;
\end{lstlisting}
\vspace{0.2cm}

The use of the C++ arithmetic operators allows for a very concise description of the addition
operation: The two vectors \lstinline|a| and \lstinline|b| are added and the result is assigned
to the third vector \lstinline|c|. Assuming that the \lstinline|Vector| class allows access to
its elements via the subscript operator and provides a \lstinline|size| function to query its
current size, \lstinline|operator+| is usually implemented similar to the following code:

\begin{lstlisting}[numbers=left,
                   frame=tb,
                   caption={[Classic implementation of the addition operator.]
                             Classic implementation of the addition operator.},
                   label={lst:idea_2}]
inline const Vector operator+( const Vector& a, const Vector& b )
{
   Vector tmp( a.size() );

   for( size_t i=0; i<a.size(); ++i )
      tmp[i] = a[i] + b[i];

   return tmp;
}
\end{lstlisting}
\vspace{0.2cm}

Although very intuitive to use and very flexible (it is for instance possible to concatenate vector
additions), the performance of this implementation in comparison with hand-crafted C code exhibits
bad performance due to the creation of the temporary vector \lstinline|tmp| in line 6. The creation
of \lstinline|tmp| involves a dynamic memory allocation, a copy operation from the temporary into
the target vector, and a memory deallocation. Additionally, the temporary interferes with cache
locality due to the increased memory footprint of the operation. All this additional overhead,
however, could be removed by implementing the vector addition manually:

\begin{lstlisting}[numbers=left,
                   frame=tb,
                   caption={[C-like, manual implementation of the addition of two vectors.]
                             C-like, manual implementation of the addition of two vectors.},
                   label={lst:idea_3}]
for( size_t i=0; i<size; ++i )
   c[i] = a[i] + b[i];
\end{lstlisting}
\vspace{0.2cm}

The performance loss is even worse if several vectors are added within a single statement due to
the ``greedy'' expression evaluation~\cite{Abrahams:2005:C++TMP}:

\begin{lstlisting}[numbers=left,
                   frame=tb,
                   caption={[Addition of three dense vectors.]
                             Addition of three dense vectors.},
                   label={lst:idea_4}]
Vector a, b, c, d;
// ... Initialization of vector a, b, and c
d = a + b + c;
\end{lstlisting}
\vspace{0.2cm}

For each single addition operation a separate temporary vector is created, whereas the operation
would not require a single temporary:

\begin{lstlisting}[numbers=left,
                   frame=tb,
                   caption={[C-like, manual implementation of the addition of three vectors.]
                             C-like, manual implementation of the addition of three vectors.},
                   label={lst:idea_5}]
for( size_t i=0; i<size; ++i )
   d[i] = a[i] + b[i] + c[i];
\end{lstlisting}
\vspace{0.2cm}

The ET approach is to create a compile-time parse tree of the whole expression to
remove the creation of the costly temporary objects entirely and to delay the execution
of the expression until it is assigned to its target. Therefore the addition operator no
longer returns the (computationally expensive) result of the addition, but a small temporary
object that acts as a placeholder for the addition expression~\cite{haerdtlein:2007:Diss}:

\begin{lstlisting}[numbers=left,
                   frame=tb,
                   caption={[ET-based implementation of the addition operator.]
                             ET-based implementation of the addition operator.},
                   label={lst:idea_6}]
template< typename A, typename B >
class Sum
{
 public:
  explicit Sum( const A& a, const B& b )
    : a_( a )
    , b_( b )
  {}

  std::size_t size() const {
    return a_.size();
  }

  double operator[]( std::size_t i ) const {
    return a_[i] + b_[i];
  }

 private:
  const A& a_;  // Reference to the left-hand side operand
  const B& b_;  // Reference to the right-hand side operand
};


template< typename A, typename B >
Sum<A,B> operator+( const A& a, const B& b )
{
  return Sum<A,B>( a, b );
}
\end{lstlisting}
\vspace{0.2cm}

Instead of calculating the result of the addition of two vectors, the addition operator now
returns an object of type \lstinline|Sum<A,B>|, where \lstinline|A| and \lstinline|B| are the
types of the left- and right-hand side operands, respectively. The only requirements the
addition operator poses on \lstinline|A| and \lstinline|B| are the existence of a subscript
operator to access the elements of the operands and a \lstinline|size| function. The
\lstinline|Sum| class has two data members, which are references-to-const to the two
operands of the addition operation. Therefore this object is cheap to create and copy
in comparison to the complete result vector. Since the \lstinline|Sum| class represents the
result of an addition, it must provide access to the resulting elements. For this purpose,
it defines two access functions: The \lstinline|size| function to access the size of the
resulting vector and the subscript operator to access the individual elements.

The \lstinline|Sum| class now temporarily represents the addition, until a special assignment
operator is encountered:

\begin{lstlisting}[numbers=left,
                   frame=tb,
                   caption={[Implementation of the ET assignment operator.]
                             Implementation of the ET assignment operator.},
                   label={lst:idea_7}]
class Vector
{
 public:
  // ...

  template< typename A >
  Vector& operator=( const A& expr )
  {
    resize( expr.size() );

    for( std::size_t i=0; i<expr.size(); ++i )
      v_[i] = expr[i];

    return *this;
  }

  // ...
};
\end{lstlisting}
\vspace{0.2cm}

This assignment operator is the only other assignment operator of the \lstinline|Vector| class
next to the copy assignment operator (which is necessary in case of a manual management of the
memory for the vector elements). Every time an expression object is assigned to a \lstinline|Vector|,
this assignment operator is used to handle the assignment\footnote{The thorough reader might
notice that due to the signature of this assignment operator all non-vector objects assigned
to a vector that do not fit the signature of the copy assignment operator will use this
assignment operator. How this problem is handled is explained in detail
in~\cite{haerdtlein:2007:Diss} and~\cite{iglberger:phd:10}.}. It first resizes the vector
accordingly and afterwards traverses the elements of the given expression within a single
\lstinline|for|-loop. Note that during this traversal the evaluation of the expression is
triggered due to the access to the values via the subscript operator. Also note that this
\lstinline|for|-loop is the only \lstinline|for|-loop necessary to evaluate the entire
expression.

Via this formulation based on the inline formulation of all functions and the evaluation within
a single \lstinline|for|-loop hidden in the assignment operator the compiler is able to generate
code similar to a C-like implementation (see Listing~\ref{lst:idea_3}). It is even possible to
concatenate several additions as for instance illustrated in Listing~\ref{lst:idea_4} without
the creation of any temporary object (and still a single \lstinline|for|-loop evaluation as in
Listing~\ref{lst:idea_5}).


%

Both the Boost \ublas{} as well as the \blitz{} library are based on the two major ideas of the
illustrated ET implementation:

\begin{itemize}
   \item no temporaries are created during the evaluation of an expression (except for the
      ET objects themselves, which also have to be considered temporaries)
   \item the elements of the left-hand side target are evaluated element-wise by the time the
      assignment operator is called and by accessing the elements of the right-hand side expression
\end{itemize}

In the following, we are comparing the performance of six different implementation of the
addition of two dense vectors. The first contestant is the classic C++ operator overloading
technique. Contestant number two is a C-like, manual implementation of the \lstinline|for|-loop,
as illustrated in Listing~\ref{lst:idea_3}. The third approach is a plain function that accepts
the two operands and the target vector of type \lstinline|Vector| as arguments and wraps the vector
addition:

\begin{lstlisting}[numbers=left,
                   frame=tb,
                   caption={[Implementation of the addition of two vectors in a plain function.]
                             Implementation of the addition of two vectors in a plain function.},
                   label={lst:idea_9}]
inline void addVectors( const Vector& a, const Vector& b, Vector& c )
{
   // ... Same implementation as in Listing 2, except no temporary is created
}
\end{lstlisting}
\vspace{0.2cm}

Contestants four and five are the \blitz{} and Boost \ublas{} libraries, respectively.
The sixth contestant is the \blaze{} library that will be introduced in
Section~\ref{sec:smart_et}.

\begin{figure}[h!t]
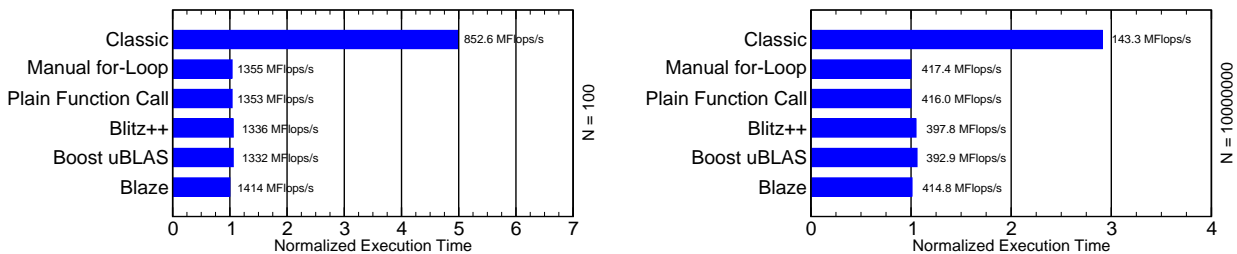

   \centering
   \includegraphics[clip=true,trim=0.5mm 0.9mm 0mm 0mm,angle=270,scale=0.29]{dvecdvecadd_100.eps}
   \hfill
   \includegraphics[clip=true,trim=0.5mm 0.9mm 0mm 0mm,angle=270,scale=0.29]{dvecdvecadd_10000000.eps}
   \caption[Performance comparison of dense vector addition implementations]{
      \label{fig:dvecdvecadd}
      Performance comparison between six different implementations of the addition between two
      dense vectors.
   }
\end{figure}

Figure~\ref{fig:dvecdvecadd} shows the performance results for both small vectors (in-cache) and
large vectors (out-of-cache). As expected, the classic C++ operator overloading shows by far the
worst performance due to the extra data transfer caused by the temporary vector. In this direct
comparison it becomes obvious that the overhead due to the creation of a temporary vector prevents
good performance. In this regard, ETs can be considered a performance optimization in
comparison to naive C++ operator overloading: By avoiding the creation of an intermediate
temporary, they achieve the performance of a manual, C-like implementation of the vector addition.
Additionally, they provide the expressiveness, naturalness, and flexibility of a domain specific
language~\cite{Abrahams:2005:C++TMP} by exploiting operator overloading, i.e. it is for instance
possible to intuitively concatenate the addition of several vectors.


\section{ETs: A Performance Optimization Technique?}
\label{sec:mmm}

The reputation that ETs are a performance optimization exclusively results from their performance
advantage compared to classic C++ operator overloading in BLAS level 1 operations, such as
the dense vector addition. No further performance comparisons have been published so far.
One main reason for that is that the optimization of array operations is, according to Veldhuizen's
original publication, the main application of ETs. Still, both \blitz{} as well as Boost \ublas{}
provide functionality well beyond the BLAS level 1 operations, which, according to the \blitz{}
homepage, also ``[...] provides performance on par with Fortran 77/90''. In this section, we will
evaluate the performance of a BLAS level 3 function, the multiplication between two dense
matrices. The characteristics of the dense matrix multiplication make this operation a
particularly well suited candidate for optimization, since with a proper optimization (memory
access scheme, etc.) this operation can be made arithmetically bound instead of memory
bound~\cite{Hager:2011:HPCSE}.

For this comparison we use six different implementations of a plain multiplication between two
dense matrices. The first implementation is a straight forward
C++ implementation using the classic operator overloading technique. Listing~\ref{lst:mmm_1} shows
the according implementation that, except for a suited ordering of the nested \lstinline|for|-loops
does not contain any optimizations.

\begin{lstlisting}[numbers=left,
                   frame=tb,
                   caption={[Implementation of the matrix-matrix multiplication operator.]
                             Implementation of the matrix-matrix multiplication operator.},
                   label={lst:mmm_1}]
inline const Matrix operator*( const Matrix& A, const Matrix& B )
{
   Matrix C( A.rows(), B.columns() );

   for( size_t i=0; i<A.rows(); ++i ) {
      for( size_t k=0; k<B.columns(); ++k ) {
         C(i,k) = A(i,0) * B(0,k);
      }
      for( size_t j=1; j<A.columns(); ++j ) {
         for( size_t k=0; k<B.columns(); ++k ) {
            C(i,k) += A(i,j) * B(j,k);
         }
      }
   }

   return C;
}
\end{lstlisting}
\vspace{0.2cm}

The second contestant is the implementation of a plain function accepting the three involved
matrices as arguments. This function is similar to the \lstinline|addVector| function from
Listing~\ref{lst:idea_9}. The third and fourth contestant are the \blitz{} (see
Listing~\ref{lst:mmm_3}) and Boost \ublas{} (see Listing~\ref{lst:mmm_4}) libraries, respectively.
The fifth implementation is provided by the \blaze{} library (see Listing~\ref{lst:mmm_5})
and the sixth code uses a plain call to the \lstinline|dgemm| BLAS function.

%

\begin{lstlisting}[numbers=left,
                   frame=tb,
                   caption={[Use of the matrix multiplication in the \blitz{} library.]
                             Use of the matrix multiplication in the \blitz{} library.},
                   label={lst:mmm_3}]
blitz::Array<double,2> A( N, N ), B( N, N ), C( N, N );
blitz::firstIndex i;
blitz::secondIndex j;
blitz::thirdIndex k;
// ... Initialization of the matrices
C = blitz::sum( A(i,k) * B(k,j), k );
\end{lstlisting}
\vspace{0.2cm}

\begin{lstlisting}[numbers=left,
                   frame=tb,
                   caption={[Use of the matrix multiplication in the Boost \ublas{} library.]
                             Use of the matrix multiplication in the Boost \ublas{} library.},
                   label={lst:mmm_4}]
boost::numeric::ublas::matrix<double> A( N, N ), B( N, N ), C( N, N );
// ... Initialization of the matrices
noalias( C ) = prod( A, B );
\end{lstlisting}
\vspace{0.2cm}

\begin{lstlisting}[numbers=left,
                   frame=tb,
                   caption={[Use of the matrix multiplication in the \blaze{} library.]
                             Use of the matrix multiplication in the \blaze{} library.},
                   label={lst:mmm_5}]
pe::MatN A( N, N ), B( N, N ), C( N, N );
// ... Initialization of the matrices
C = A * B;
\end{lstlisting}
\vspace{0.2cm}


\begin{figure}[h!t]
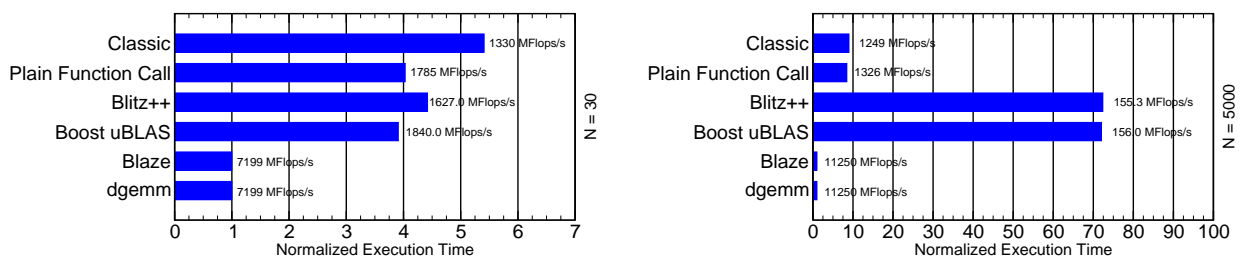

   \centering
   \includegraphics[clip=true,trim=0.5mm 0.9mm 0mm 0mm,angle=270,scale=0.29]{dmatdmatmult_30.eps}
   \hfill
   \includegraphics[clip=true,trim=0.5mm 0.9mm 0mm 0mm,angle=270,scale=0.29]{dmatdmatmult_5000.eps}
   \caption[Performance comparison of dense matrix multiplication implementations]{
      \label{fig:dmatdmatmult}
      Performance comparison between five different implementations of the multiplication between
      two dense matrices.
   }
\end{figure}

Figure~\ref{fig:dmatdmatmult} shows the performance results for the six different implementations.
For both an in-cache matrix multiplication with two matrices of size $30^2$ as well as the
out-of-cache multiplication with two matrices of size $5000^2$ the \lstinline|dgemm| function
is clearly the fastest competitor. However, although also based on ETs, the \blaze{} library
achieves the same performance level, since internally the \blaze{} also uses the \lstinline|dgemm|
function (see Section~\ref{sec:smart_et}). In contrast, the other two ET-based libraries exhibit
very poor performance. Whereas this result comes with no real surprise, since the purpose of the
\lstinline|dgemm| function is to provide a maximum of performance for the matrix multiplication,
the fact that even the simple, non-optimized, old fashioned operator overloading performs much
better in case of the out-of-cache matrices than the ET-based libraries is surprising.

\begin{table}
\begin{center}
\begin{footnotesize}
\begin{tabular}{ll|p{2.1cm}|p{2.1cm}|p{2.1cm}|p{2.1cm}p{0.01cm}}
\multicolumn{2}{c|}{}                                                                          & \centering Memory Bandwidth [MBytes/s] & \centering Total Retired Instructions [$10^{11}$] & \centering Total Arithmetic Operations [$10^{11}$] & \centering Cycles Per Instruction (CPI) & \\[1mm]
\hline
                                                                         & STREAM              & \centering 11814                       & \centering ---                                    & \centering ---                                     & \centering ---        & \\[1mm]
\hline
\multirow{6}{*}{\begin{sideways}\centering N=5000 $\quad$\end{sideways}} & Classic             & \centering  5008                       & \centering $12.5054$                              & \centering $2.50231$                               & \centering $0.441127$ & \\[1mm]
                                                                         & Plain Function Call & \centering  5328                       & \centering $12.5048$                              & \centering $2.50232$                               & \centering $0.440912$ & \\[1mm]
                                                                         & \blitz              & \centering   623                       & \centering $10.0126$                              & \centering $2.58185$                               & \centering $4.67952$  & \\[1mm]
                                                                         & Boost \ublas        & \centering   623                       & \centering $10.0053$                              & \centering $2.50197$                               & \centering $4.72096$  & \\[1mm]
                                                                         & \blaze              & \centering   496                       & \centering $ 2.02589$                             & \centering $2.50612$                               & \centering $0.322074$ & \\[1mm]
                                                                         & dgemm               & \centering   496                       & \centering $ 2.02589$                             & \centering $2.50612$                               & \centering $0.322074$ & \\[1mm]
\end{tabular}
\end{footnotesize}
\end{center}
\caption[Likwid performance analysis of the multiplication between two dense matrices]{
   \label{tab:dmatdmatmult}
   Likwid performance analysis of the multiplication between two dense matrices. Note that the
   \lstinline|dgemm| function uses packed instructions, which may results in a higher number
   of arithmetic operations than retired instructions!
}
\end{table}

Table~\ref{tab:dmatdmatmult} gives an indication of why the performance of Boost \ublas{} and
\blitz{} is so bad. We used the Likwid tool suite~\cite{treibig:10} to measure the achieved
memory bandwidth, the total number of retired instructions, the total number of arithmetic
operations, and the number of cycles per instruction (CPI). When comparing the CPI both
Boost \ublas{} as well as \blitz{} reveal a low quality of the generated code. In combination
with the high number of retired instructions and the low achieved memory bandwidth it
becomes obvious why the used ET implementation is shows low performance.

The reason for this behavior is intrinsic to the methodology of ETs. Based on the philosophy
that each element of the target data structure is computed one after another, the executed
code is similar to the code shown in Listing~\ref{lst:mmm_7}:

\begin{lstlisting}[numbers=left,
                   frame=tb,
                   caption={[Slow implementation of the matrix-matrix multiplication operator.]
                             Slow implementation of the matrix-matrix multiplication operator.},
                   label={lst:mmm_7}]
for( size_t i=0; i<A.rows(); ++i ) {
   for( size_t j=0; j<B.columns(); ++j ) {
      for( size_t k=0; k<A.columns(); ++k ) {
         C(i,j) += A(i,k) * B(k,j);
      }
   }
}
\end{lstlisting}
\vspace{0.2cm}

This loop ordering corresponds to the worst possible data access scheme that can be used for the
matrix multiplication: For each element of the target matrix a complete column of the right-hand
side matrix is traversed, resulting in a cache line transfer for each individual data value.
Especially for out-of-cache matrices, this approach is very cache inefficient since only
a single value of each cache line can be used before the cache line has to be replaced. In
contrast to this loop ordering, the two codes for classic operator overloading and the plain
function call use a more cache efficient data access scheme that simultaneously calculates
several values of the target matrix, which results in a much better memory bandwidth and
lower CPI.

Although in case of the classic operator overloading technique a temporary is created, which
is omitted in case of the ET libraries, the performance of the classic technique is much
better. Obviously, the performance gain results from the choice
of the better data access scheme. Thus the primary question is why the ET libraries don't
implement the more efficient loop ordering in order to gain performance. The reason for this is
that with the current methodology ETs are not able to choose the best data access scheme. ETs
are solely based on the goal to avoid temporaries, the strategy to evaluate the given
right-hand side expressions element-wise, and the firm believe that the compiler will optimize
the resulting code constructs after inlining took place. Whereas this works well for array
operations as for instance the vector addition, where there is barely opportunity for data
access scheme optimization and where therefore the omission of the temporary results in a
performance optimization, in order to achieve high performance for the matrix multiplication
the detailed knowledge about the involved data structures and the operation has to be exploited.

The fundamental problem of the current ET technique is that it is no performance optimization
technique, but essentially an abstraction technique. Whereas this abstraction improves the
flexibility of a framework to integrate new types and operations, it counteracts high performance
on several levels. First, ETs abstract from the involved data
types. A clear indication for this is that the involved ET data types are required to adhere to
a certain interface (``Design by Contract''~\cite{meyer:97}). Therefore no special optimization
can be applied based on the type of the used matrices. Second, ETs abstract from the type of
operation. From an abstract point of view it makes no difference whether the target matrix is
assigned a matrix addition expression or a matrix multiplication expression; In both cases, the
according assignment operator accesses the elements of this virtual matrix to fill the target
matrix. However, in terms of performance a matrix addition has to be treated fundamentally
different from a matrix multiplication. Therefore, with the current methodology, real
performance optimization based on memory optimization (the most important optimization for
contemporary, cache-based architectures~\cite{Hager:2011:HPCSE}), vectorization, and
exploitation of superscalarity, cannot be properly performed. The optimization capability
of ETs is thus limited to operations where the abstract data access scheme coincidentally
corresponds to the optimal data access scheme.

These results have another important implication. A crucial aspect of ETs is the encapsulation
of the numerical operations in functions. By this they provide an intuitive, easy-to-use interface
and a very high maintainability. This aspect is especially important for complex numerical
operations, such as the matrix multiplication: Whereas simple numerical kernels, such as a
vector addition, can easily be rewritten, it should not be necessary to repeatedly reimplement
complex kernels, which contain a huge amount of work to achieve high performance. From a
performance point of view, the encapsulation of complex kernels is therefore more important
than the encapsulation of simple kernels. Considering the performance results for the matrix
multiplication, it must be concluded that the current ET methodology is not suitable to
encapsulate highly optimized complex kernels.

\section{Sparse Arithmetic}
\label{sec:sparse}


Due to the abstraction from the actual data types in all operations, ETs offer an impressive
flexibility to integrate new data types into the system. The abstraction is achieved by
requiring all data types to adhere to a certain interface via which it is possible to access
the underlying elements. One example, for what this flexibility can be used, is demonstrated
by the Boost \ublas{} library: In contrast to \blitz, Boost \ublas{} provides sparse vectors and
matrices that can be homogeneously combined with the available dense vectors and matrices.
This enriched functionality is clearly an extraordinary strength of ETs. The downside of
this abstraction, however, is a performance penalty. In order to show this performance
penalty, we selected two operations between dense and sparse data types and compared their
performance between Boost \ublas{} and the \blaze.

The first operation is the multiplication between a row-wise stored sparse matrix and a dense vector.
This type of operation is of importance in many engineering applications as it is for instance
used to solve linear systems of equations. Listing~\ref{lst:sparse_1} shows its implementation
with the Boost \ublas{} library, Listing~\ref{lst:sparse_2} with the \blaze{} library.

\begin{lstlisting}[numbers=left,
                   frame=tb,
                   caption={[Use of the sparse matrix/dense vector multiplication in the Boost \ublas{} library.]
                             Use of the sparse matrix/dense vector multiplication in the Boost \ublas{} library.},
                   label={lst:sparse_1}]
boost::numeric::ublas::compressed_matrix<double> A( N, N );
boost::numeric::ublas::vector<double> a( N ), b( N );
// ... Initialization of the matrix and the vectors
noalias( b ) = prod( A, a );
\end{lstlisting}
\vspace{0.2cm}

\newpage

\begin{lstlisting}[numbers=left,
                   frame=tb,
                   caption={[Use of the sparse matrix/dense vector multiplication in the \blaze{} library.]
                             Use of the sparse matrix/dense vector multiplication in the \blaze{} library.},
                   label={lst:sparse_2}]
SparseMatrixMxN<double> A( N, N );
Vector<double> a( N ), b( N );
// ... Initialization of the matrix and the vectors
b = A * a;
\end{lstlisting}
\vspace{0.2cm}

\begin{figure}[h!t]
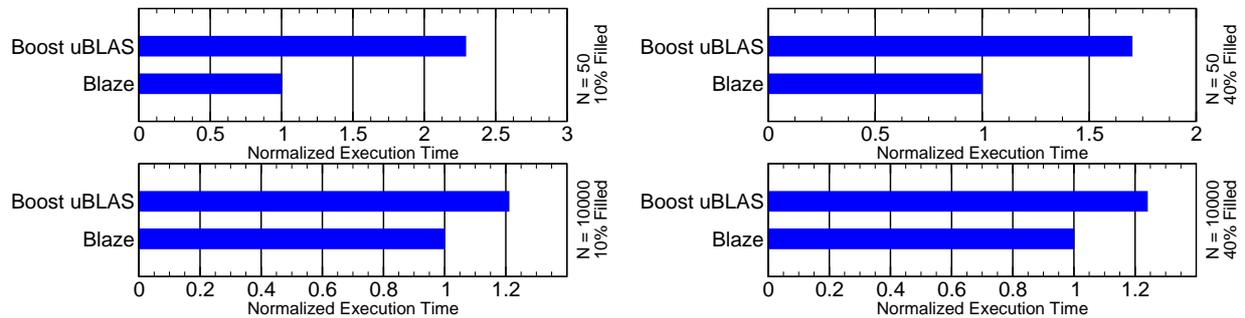

   \centering
   \includegraphics[clip=true,trim=0.5mm 0.9mm 0mm 0mm,angle=270,scale=0.31]{smatdvecmult_50_10.eps}
   \hfill
   \includegraphics[clip=true,trim=0.5mm 0.9mm 0mm 0mm,angle=270,scale=0.31]{smatdvecmult_50_40.eps}
   \includegraphics[clip=true,trim=0.5mm 0.9mm 0mm 0mm,angle=270,scale=0.31]{smatdvecmult_10000_10.eps}
   \hfill
   \includegraphics[clip=true,trim=0.5mm 0.9mm 0mm 0mm,angle=270,scale=0.31]{smatdvecmult_10000_40.eps}
   \caption[Performance comparison of sparse matrix/dense vector multiplication implementations]{
      \label{fig:smatdvecmult}
      Performance comparison between Boost \ublas{} and \blaze{} for the multiplication between
      a sparse matrix and a dense vector.
   }
\end{figure}

Figure~\ref{fig:smatdvecmult} shows the in-cache and out-of-cache performance results for a 10\%
and 40\% filled sparse matrix, respectively. The direct comparison between Boost \ublas{} and the
\blaze{} does not exhibit a huge performance difference neither for the different sizes nor the
different filling degrees. The reason for that is that the default memory access scheme utilized
by the ET implementations works perfectly for this operation: A single row of the matrix has to
be multiplied with the dense vector for each result vector element. Since both the row-wise
memory access to the sparse matrix as well as the access to the dense vector perfectly exploit
the structure of both data structures, the performance is on a reasonable level.

\begin{lstlisting}[numbers=left,
                   frame=tb,
                   caption={[Use of the dense matrix/sparse matrix multiplication in the Boost \ublas{} library.]
                             Use of the dense matrix/sparse matrix multiplication in the Boost \ublas{} library.},
                   label={lst:sparse_3}]
boost::numeric::ublas::matrix<double> A( N, N ), C( N, N );
boost::numeric::ublas::compressed_matrix<double> B( N, N );
// ... Initialization of the matrix and the vectors
noalias( C ) = prod( A, B );
\end{lstlisting}
\vspace{0.2cm}

\begin{lstlisting}[numbers=left,
                   frame=tb,
                   caption={[Use of the dense matrix/sparse matrix multiplication in the \blaze{} library.]
                             Use of the dense matrix/sparse matrix multiplication in the \blaze{} library.},
                   label={lst:sparse_4}]
MatrixMxN<double> A( N, N ), C( N, N );
SparseMatrixMxN<double> B( N, N );
// ... Initialization of the matrices
C = A * B;
\end{lstlisting}
\vspace{0.2cm}

\begin{figure}[h!t]
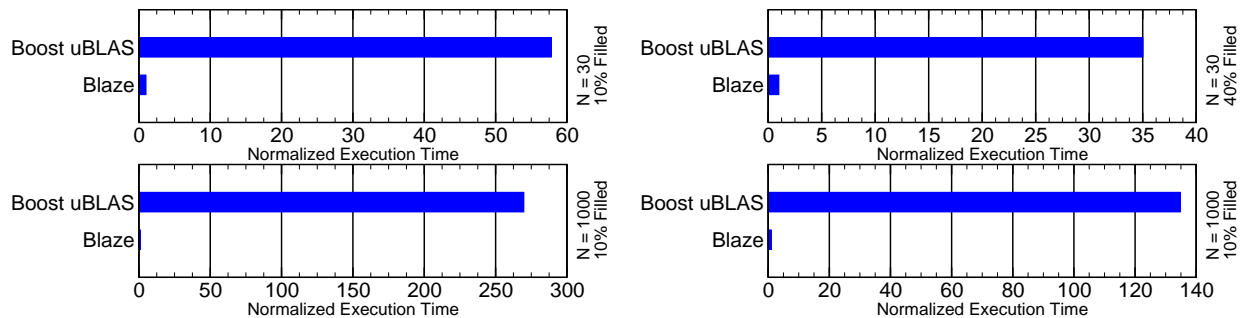

   \centering
   \includegraphics[clip=true,trim=0.5mm 0.9mm 0mm 0mm,angle=270,scale=0.31]{dmatsmatmult_30_10.eps}
   \hfill
   \includegraphics[clip=true,trim=0.5mm 0.9mm 0mm 0mm,angle=270,scale=0.31]{dmatsmatmult_30_40.eps}
   \includegraphics[clip=true,trim=0.5mm 0.9mm 0mm 0mm,angle=270,scale=0.31]{dmatsmatmult_1000_10.eps}
   \hfill
   \includegraphics[clip=true,trim=0.5mm 0.9mm 0mm 0mm,angle=270,scale=0.31]{dmatsmatmult_1000_40.eps}
   \caption[Performance comparison of dense matrix/sparse matrix multiplication implementations]{
      \label{fig:dmatsmatmult}
      Performance comparison between Boost \ublas{} and \blaze{} for the multiplication between
      a dense and a sparse matrix.
   }
\end{figure}

The situation changes entirely when we multiply a row-wise stored dense matrix with a row-wise
stored sparse matrix. Listing~\ref{lst:sparse_3} shows the implementation of this operation with
the Boost \ublas{} library, Listing~\ref{lst:sparse_4} shows its implementation with the \blaze.
Figure~\ref{fig:dmatsmatmult} shows the in-cache and out-of-cache performance results for 10\%
and 40\% filled sparse matrices, respectively. It becomes obvious that there is a tremendous
performance difference between the two libraries that cannot be explained by simple differences
in the implementation of the codes, but points at fundamental differences in the methodology of
the two ET libraries. Whereas the \blaze{} attempts to exploit all information about the operations
and both data types and therefore deals as efficiently as possible with the fact that the
right-hand side sparse matrix is stored in a row-wise fashion, Boost \ublas{} completely abstracts
from the current operation and the data types of the two involved matrices. All elements of the
result matrix are evaluated one after another~\footnote{As a reminder: This approach is necessary
due to the abstraction from the actual operation!} by traversing the left-hand side dense matrix
via row-iterators and the right-hand side sparse matrix via column iterators. Although the column
iterators can be considered a very convenient interface for users of the library, their internal,
abstract use results in a devastating performance penalty in this case. What would be required
instead in order to achieve high performance would be a recognition of the data structure of the
right-hand side sparse matrix and the attempt to use and reuse its elements in a cache-efficient
manner. However, due to the abstraction from both the actual operation as well as the data types,
this is not possible. Therefore the current methodology of ETs prohibits any possible performance
optimization for this operation.

Note that this operation was specifically selected to demonstrate that performance greatly suffers
from the abstraction from the data types and operations. The performance penalty would be much less
severe in case of a column-wise stored sparse matrix. However, since ET libraries are
usually provided as black box systems, the knowledge that the combination of certain data structures
should be (completely) avoided, cannot be expected from a user of the library.


\section{Complex Expressions}
\label{sec:complex}

One of the two fundamental rules of ETs is that no temporaries are created during the evaluation
of an expression in order to avoid all overhead involved in creating a temporary. This rule is
mainly responsible for the reputation that ETs are a performance optimization. However, in
certain situations the creation of a temporary is strictly necessary to achieve performance
although it involves extra work. In this section we have specifically selected two examples
for complex expressions that require the creation of temporaries in order to demonstrate the
shortcoming of this rule.


The first complex expression is the multiplication between a dense matrix and the sum of
three dense vectors: $A \cdot ( a + b + c )$. The problem that is involved in this expression
is obvious: The right-hand side vector of the matrix-vector multiplication is required several
times during its evaluation. In case the result of the vector additions $a + b + c$ is not
computed prior to the multiplication, the additions have to be evaluated several times, which
will inevitably result in a performance loss.

\begin{lstlisting}[numbers=left,
                   frame=tb,
                   caption={[Use of the expression $d = A * ( a + b + c )$ with classic operator overloading.]
                             Use of the expression $d = A * ( a + b + c )$ with classic operator overloading.},
                   label={lst:complex_1}]
classic::Matrix<double> A( N, N );
classic::Vector<double> a( N ), b( N ), c( N ), d( N );
// ... Initialization of the matrix and vectors
d = A * ( a + b + c );
\end{lstlisting}
\vspace{0.2cm}

\begin{lstlisting}[numbers=left,
                   frame=tb,
                   caption={[Use of the expression $d = A * ( a + b + c )$ in the \blitz{} library.]
                             Use of the expression $d = A * ( a + b + c )$ in the \blitz{} library.},
                   label={lst:complex_2}]
blitz::Array<real,2> A( N, N );
blitz::Array<real,1> a( N ), b( N ), c( N ), d( N ), tmp( N );
blitz::firstIndex i;
blitz::secondIndex j;
// ... Initialization of the matrix and vectors
tmp = a + b + c;
d   = blitz::sum( A(i,j) * tmp(j), j );
\end{lstlisting}
\vspace{0.2cm}

\begin{lstlisting}[numbers=left,
                   frame=tb,
                   caption={[Use of the expression $d = A * ( a + b + c )$ in the Boost \ublas{} library.]
                             Use of the expression $d = A * ( a + b + c )$ in the Boost \ublas{} library.},
                   label={lst:complex_3}]
boost::numeric::ublas::matrix<real> A( N, N );
boost::numeric::ublas::vector<real> a( N ), b( N ), c( N ), d( N );
// ... Initialization of the matrices
noalias( d ) = prod( A, ( a + b + c ) );
\end{lstlisting}
\newpage

\begin{lstlisting}[numbers=left,
                   frame=tb,
                   caption={[Use of the expression $d = A * ( a + b + c )$ in the \blaze{} library.]
                             Use of the expression $d = A * ( a + b + c )$ in the \blaze{} library.},
                   label={lst:complex_4}]
pe::MatrixMxN<double> A( N, N );
pe::VectorN<double> a( N ), b( N ), c( N ), d( N );
// ... Initialization of the matrices
d = A * ( a + b + c );
\end{lstlisting}
\vspace{0.2cm}

\begin{figure}[h!t]
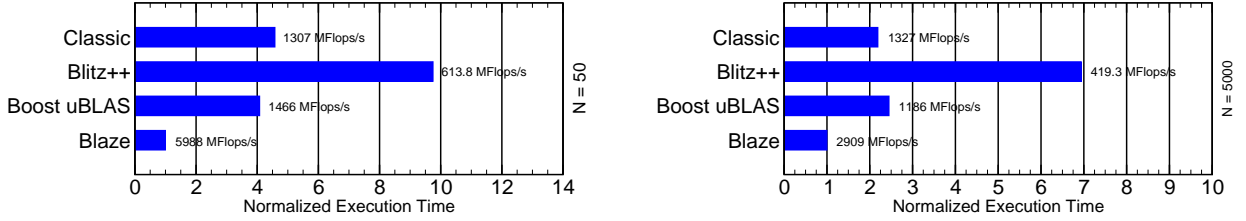

   \centering
   \includegraphics[clip=true,trim=0.5mm 0.9mm 0mm 0mm,angle=270,scale=0.31]{complex2_50.eps}
   \hfill
   \includegraphics[clip=true,trim=0.5mm 0.9mm 0mm 0mm,angle=270,scale=0.31]{complex2_5000.eps}
   \caption[Performance comparison for the complex expression $A \cdot ( a + b + c )$]{
      \label{fig:complex2}
      Performance comparison between four different implementations of the complex expression
      $A \cdot ( a + b + c )$.
   }
\end{figure}

Listings~\ref{lst:complex_1}, \ref{lst:complex_2}, \ref{lst:complex_3}, and \ref{lst:complex_4}
show the implementation of the complex expression with classic operator overloading, \blitz,
Boost \ublas{} and \blaze, respectively. Interestingly, it is necessary to explicitly create the
temporary \lstinline|tmp| in case of \blitz{} since it is syntactically not possible to evaluate
the complex expression within a single statement. Figure~\ref{fig:complex2} shows the in-case and
out-of-cache performance results of four different implementations of this expression: classic
operator overloading, the \blitz{} library, Boost \ublas, and the \blaze{} library.

For both small and large $N$, the two traditional, ET-based libraries do not exhibit good
performance. Especially in case of large $N$, classic operator overloading, although
requiring a total of three temporaries for the evaluation of the expression, performs better
than Boost \ublas{} and especially \blitz. The \blaze{} library, which uses a single temporary
to store the intermediate result of the vector additions and utilizes the optimized
\lstinline|dgemv| function for the subsequent matrix-vector multiplication, has a clear
performance advantage. Table~\ref{tab:complex2} shows the Likwid results, which allow to
analyze the performance results in more detail. It becomes obvious that the better cache
utilization as well as the memory bandwidth of the optimized \lstinline|dgemv| function
result in higher performance.

\begin{table}[h!t]
\begin{center}
\begin{footnotesize}
\begin{tabular}{ll|p{1.9cm}|p{1.9cm}|p{1.9cm}|p{1.9cm}|p{2.2cm}p{0.01cm}}
\multicolumn{2}{c|}{}                                                                   & \centering Memory Bandwidth [MBytes/s] & \centering Total Retired Instructions [$10^8$] & \centering Total Arithmetic Operations [$10^7$] & \centering Cycles Per Instruction (CPI) & \centering L1 Data Cache Line Replacements [$10^6$] & \\[1mm]
\hline
                                                                         & STREAM       & \centering 11814                       & \centering ---                                 & \centering ---                                  & \centering ---                          & \centering ---                                      & \\[1mm]
\hline
\multirow{4}{*}{\begin{sideways}\centering N=5000 $\quad$\end{sideways}} & Classic      & \centering  5387                       & \centering $4.31892$                           & \centering $ 7.56438$                           & \centering $0.455758$                   & \centering $ 6.32184$                               & \\[1mm]
                                                                         & \blitz       & \centering  2295                       & \centering $6.87758$                           & \centering $ 7.55893$                           & \centering $0.531862$                   & \centering $ 6.36924$                               & \\[1mm]
                                                                         & Boost \ublas & \centering  4382                       & \centering $4.5681$                            & \centering $12.5684$                            & \centering $0.529004$                   & \centering $12.5812$                                & \\[1mm]
                                                                         & \blaze       & \centering 11088                       & \centering $2.74818$                           & \centering $ 7.74858$                           & \centering $0.57686$                    & \centering $ 3.99694$                               & \\[1mm]
\end{tabular}
\end{footnotesize}
\end{center}
\caption[Likwid performance analysis of the complex expression $A \cdot ( a + b + c )$]{
   \label{tab:complex2}
   Likwid performance analysis of the complex expression $A \cdot ( a + b + c )$.
}
\end{table}

\begin{figure}[h!t]
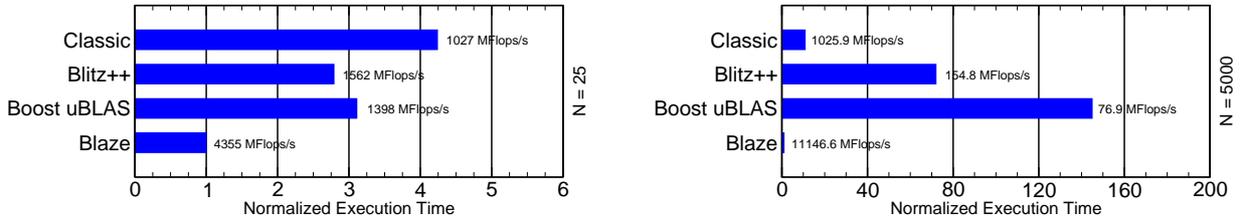

   \centering
   \includegraphics[clip=true,trim=0.5mm 0.9mm 0mm 0mm,angle=270,scale=0.31]{complex6_25.eps}
   \hfill
   \includegraphics[clip=true,trim=0.5mm 0.9mm 0mm 0mm,angle=270,scale=0.31]{complex6_5000.eps}
   \caption[Performance comparison for the complex expression $( A + B ) \cdot ( C - D )$]{
      \label{fig:complex6}
      Performance comparison between four different implementations of the complex expression
      $( A + B ) \cdot ( C - D )$.
   }
\end{figure}

The second complex expression we selected involves four dense matrices:
$E = ( A + B ) \cdot ( C - D )$. In this case, in order to efficiently be able to evaluate the
matrix multiplication, both the left-hand side as well as the right-hand side matrix expression
must be evaluated prior to the matrix multiplication.
Again, in case of \blitz, the expression cannot be computed within a single statement, which
results in two explicit temporary matrices. The benefit of this can be
seen in Figure~\ref{fig:complex6}, which shows both the in-cache as well as the out-of-cache
results for classic operator overloading, \blitz, Boost \ublas, and the \blaze{} library. \blitz{}
always performs better than Boost \ublas, which does not create any intermediate temporaries
and re-evaluates the matrix addition and subtraction repeatedly. However, both \blitz{} as well
as Boost \ublas{} exhibit poor performance in comparison to the \blaze{} library, which internally
creates two temporaries to hold the intermediate results of the matrix addition and subtraction
and uses the \lstinline|dgemm| function to compute the subsequent matrix multiplication.
Especially striking is the fact that for large $N$ both \blitz{} and Boost \ublas{} are extremely
by classic operator overloading, since it does create the necessary temporaries
and utilizes a faster kernel for the matrix multiplication. These performance are confirmed
by the Likwid results in Table~\ref{tab:complex6}. Based on the large data cache replacement
rate, the large CPI and the low memory bandwidth the quality of the generated code is poor.

\begin{table}[h!t]
\begin{center}
\begin{footnotesize}
\begin{tabular}{ll|p{1.9cm}|p{1.9cm}|p{1.9cm}|p{1.9cm}|p{2.2cm}p{0.01cm}}
\multicolumn{2}{c|}{}                                                                          & \centering Memory Bandwidth [MBytes/s] & \centering Total Retired Instructions [$10^{11}$] & \centering Total Arithmetic Operations [$10^{11}$] & \centering Cycles Per Instruction (CPI) & \centering L1 Data Cache Line Replacements [$10^9$] & \\[1mm]
\hline
                                                                         & STREAM              & \centering 11814                       & \centering ---                                    & \centering ---                                     & \centering ---                          & \centering ---                                      & \\[1mm]
\hline
\multirow{4}{*}{\begin{sideways}\centering N=5000 $\quad$\end{sideways}} & Classic             & \centering  4136                       & \centering $12.5106$                              & \centering $2.50318$                               & \centering $0.442167$                   & \centering $ 31.3006$                               & \\[1mm]
                                                                         & \blitz              & \centering   624                       & \centering $10.0163$                              & \centering $2.56789$                               & \centering $4.68541$                    & \centering $266.566$                                & \\[1mm]
                                                                         & Boost \ublas        & \centering   619                       & \centering $13.7553$                              & \centering $6.15386$                               & \centering $6.90581$                    & \centering $533.411$                                & \\[1mm]
                                                                         & \blaze              & \centering   490                       & \centering $ 2.02977$                             & \centering $2.50684$                               & \centering $0.322925$                   & \centering $  2.07864$                              & \\[1mm]
\end{tabular}
\end{footnotesize}
\end{center}
\caption[Likwid performance analysis of the complex expression $( A + B ) \cdot ( C - D )$]{
   \label{tab:complex6}
   Likwid performance analysis of the complex expression $( A + B ) \cdot ( C - D )$.
}
\end{table}

\newpage

Admittedly, a simple solution to improve the performance of \blitz{} and Boost \ublas{} would be
the explicit generation of temporaries whenever necessary. This solution is also for instance
advertised on the Boost \ublas{} homepage, where the \ublas{} designers advocate the reintroduction
of temporaries as a performance remedy\footnote{In fact, the suggested solutions often involve
ungraceful syntactical expressions that don't have anything left from the elegance ETs try to
achieve}. However, arguably the primary goals of ETs is the ability to use infix operator notion
and to provide a convenient, intuitive black box interface for all kinds of mathematical operations.
Therefore a user of these libraries cannot be blamed for the lack of proper automatic recognition
of necessary temporaries. Since these libraries provide this interface, they have to expect
that someone actually uses the interface and therefore have to take care of all possible
consequences, including the requirement to create automatic temporaries. The fundamental ET
rule to avoid all temporaries, which established the reputation of ETs being a performance
optimization, can therefore obviously also act as performance ``pessimization''.



\section{A New ET Methodology: Smart Expression Templates}
\label{sec:smart_et}


Obviously ETs themselves are not generally able to provide high performance. However, the idea
to combine high performance code with the mathematical syntax provided by the C++ operators is
a justified one: code clarity, readability, and maintainability are greatly improved if used in
a mathematical context.

In this section we will shortly sketch the smart ET methodology of the \blaze{} library by means
of the multiplication between two dense matrices. In contrast to other ET frameworks the ET
implementation of the \blaze{} library takes a
completely different approach in order to achieve the goal of combining performance and syntax.
In \blaze{} the notion of ETs being a performance optimization is completely dropped. The ETs
merely act as a parsing functionality that understands the structure of the given mathematical
expression, knows the order in which subexpressions need to be evaluated (including
the creation of temporaries; see Section~\ref{sec:complex})\footnote{An illustrating example for
a smart choice for the evaluation order of subexpressions is the expression $A * B * v$, where $A$
and $B$ are two matrices and $v$ is a vector. Usually the expression is evaluated from left to
right, resulting in a matrix-matrix multiplication and a subsequent matrix-vector multiplication.
However, if the right subexpression were be evaluated first, the performance can be dramatically
improved since the matrix-matrix multiplication can be avoided in favor of a second matrix-vector
multiplication.}, and selects the appropriate, highly optimized kernels, which provide a manual,
architecture-specific performance optimization based on detailed knowledge of data types and
operations.

Note that this section serves only as a cursory overview of the methodology of smart expression
templates. We will not go into detail of the very sophisticated C++ implementation; The
implementation along with a detailed discussion will be published in a separate article. Instead,
we only try to explain the idea of the smart ET methodology by focusing on the two key concepts:
the selective creation of intermediate temporaries and the integration of optimized kernels.

\subsection{Creation of Intermediate Temporaries}
\label{subsec:temporaries}

The first key idea of smart ETs involves the creation of intermediate temporaries. The
fundamental rule of ETs not to create temporaries has two different reasons. The first
reason is the abstraction from the actual operations. Due to that the necessity to create
a temporary cannot be recognized. The second reason seems to be the apparent impossibility
to efficiently create intermediate temporaries from subexpressions inside an expression
object, since the creation of a temporary is always associated with a memory allocation, a
copy operation, and a memory deallocation.

The solution for this problem is already incorporated in the C++ standard itself. Consider
the following operation:

\begin{lstlisting}[numbers=left,
                   frame=tb,
                   caption={[Addition of three dense vectors.]
                             Addition of three dense vectors.},
                   label={lst:smart_et_1}]
Vector a, b;
// Initialization of vector a and b
Vector c = a + b;  // Same as: Vector c( a + b );
\end{lstlisting}
\vspace{0.2cm}

In contrast to the dense vector addition shown in Listing~\ref{lst:idea_1}, in this case we do
not perform an assignment, but rather an initialization of the dense vector \lstinline|c|.
Therefore there is no performance difference between all implementations: Even classic
operator overloading exhibits the same performance as the ET-based libraries. The reason behind
this is the ``named return value'' (NRV) optimization (see section 12.1.1c of the
ARM~\cite{ellis:1990:ARM} or~\cite{Lippman:2007:IC++OM}). Listing~\ref{lst:smart_et_2}
shows the compiler-optimized implementation of the addition operator from Listing~\ref{lst:idea_2}.
If the compiler applies NRV to a code (which is triggered by the presence of an explicit
copy constructor), the local variable \lstinline|tmp| will be replaced by a reference to the
eventual destination of the return value in the caller and instead of returning a temporary
the function returns \lstinline|void|.

\begin{lstlisting}[numbers=left,
                   frame=tb,
                   caption={[NRV optimization of the dense vector addition operator.]
                             NRV optimization of the dense vector addition operator.},
                   label={lst:smart_et_2}]
inline void operator+( Vector& dest, const Vector& lhs, const Vector& rhs )
{
  dest.Vector::Vector( lhs.size() );  // Explicit constructor call

  for( std::size_t i=0; i<lhs.size(); ++i )
    dest[i] = lhs[i] + rhs[i];
}
\end{lstlisting}
\vspace{0.2cm}

In case of an initialization, the compiler can therefore directly write the result to the
destination vector, which corresponds to the behavior that is achieved by the ET formulation.
In case of an assignment, a temporary is created by means of the NRV optimized code, which is
in turn assigned to the destination vector:

\begin{lstlisting}[numbers=left,
                   frame=tb,
                   caption={[Compiler generated code for the copy assignment of vectors]
                             Compiler generated code for the copy assignment of vectors},
                   label={lst:smart_et_3}]
Vector a, b, c;

// NRV optimized addition of a and b into the temporary tmp
Vector tmp( a + b );

// Assignment of the temporary to the vector c
c = tmp;
\end{lstlisting}
\vspace{0.2cm}

In the context of ETs, in case an intermediate temporary is created (for instance as the result
of a subexpression), it is created via initialization (not assignment). The same is true for the
creation of the temporary expression objects themselves. Hence the creation of temporaries
does not involve a single copy operation, but only the necessary memory allocations and
deallocations. Therefore in the smart ET methodology temporary objects are used to hold
intermediate results of subexpressions as member variables of other expression objects.

\subsection{Integration of Optimized Compute Kernels}
\label{subsec:kernels}

The second key idea of smart expression templates is the selection of the appropriate
compute kernel. The solution is to omit the abstract assignment via the assignment
operator, by passing this responsibility to the resulting expression object. Since
the expression object holds all knowledge about the involved data types and operations
it can perform the assignment as efficiently as possible. The following code excerpt
shows how this optimization is implemented in case of the \lstinline|DMatDMatMultExpr|
class that represents the multiplication between two dense matrices:

\begin{lstlisting}[numbers=left,
                   frame=tb,
                   caption={[Smart expression object for the matrix-matrix multiplication]
                             Smart expression object for the matrix-matrix multiplication},
                   label={lst:smart_et_4}]
template< typename MT1    // Type of the left-hand side dense matrix
        , typename MT2 >  // Type of the right-hand side dense matrix
class DMatDMatMultExpr : private Expression
{
 public:
  // Public interface omitted

 private:
  // ...

  // Result type of the left-hand side dense matrix expression
  typedef typename MT1::ResultType     RT1;

  // Result type of the right-hand side dense matrix expression
  typedef typename MT2::ResultType     RT2;

  // Composite type of the left-hand side dense matrix expression
  typedef typename MT1::CompositeType  CT1;

  // Composite type of the right-hand side dense matrix expression
  typedef typename MT2::CompositeType  CT2;

  // Member data type of the left-hand side dense matrix expression.
  typedef typename SelectType<IsExpression<MT1>::value,const RT1,CT1>::Type  Lhs;

  // Member data type of the right-hand side dense matrix expression.
  typedef typename SelectType<IsExpression<MT2>::value,const RT2,CT2>::Type  Rhs;

  Lhs lhs_;  // Left-hand side dense matrix of the multiplication expression.
  Rhs rhs_;  // Right-hand side dense matrix of the multiplication expression.

  // Specialized assign function injected into the surrounding namespace
  template< typename MT >  // Type of the target dense matrix
  friend inline void assign( DenseMatrix<MT>& lhs,
                             const DMatDMatMultExpr& rhs )
  {
    // Depending on the data type utilization of the cblas_dgemm kernel or
    // use of the default implementation of the matrix-matrix multiplication
  }

  // ...
};
\end{lstlisting}
\vspace{0.2cm}

The \lstinline|DMatDMatMultExpr| is implemented as a template parameterized with the two
data types \lstinline|MT1| and \lstinline|MT2| of the involved dense matrices.
The class is derived from the \lstinline|Expression| class, which makes the
\lstinline|DMatDMatMultExpr| class an expression (in contrast to plain matrices). Via
``Template Meta Programming'' (TMP)~\cite{Abrahams:2005:C++TMP} the data types of the two
operands are used to evaluate the two member data types \lstinline|Lhs| and \lstinline|Rhs|.
In case any of these types is an expression (i.e., derived from the \lstinline|Expression|
class), the \lstinline|ResultType| of the according matrix expression is used to create
a temporary object (optimized by the NRV optimization and therefore without a copy
operation). Otherwise the \lstinline|CompositeType| of the matrix expression is
used, which represents the knowledge of the expression how it should be treated in a
composite expression.

The core of the class is the \lstinline|assign| function, which implements the assignment
of the matrix-matrix multiplication to a dense matrix. This function is injected into the
surrounding namespace via the Barton-Nackman trick~\cite{barton95,vandevoorde:2003:C++}.
In case of an assignment of a temporary \lstinline|DMatDMatMultExpr| object to a
dense matrix this function is called, which performs the assignment of the
matrix multiplication based on the fastest available compute kernel. Depending on the
types of the matrix operands it either applies a default matrix multiplication kernel
(which works with any data type) or a call to the optimized BLAS functions
(\lstinline|cblas_sgemm| for single-precision matrices and \lstinline|cblas_dgemm| for
double-precision matrices).

In summary, the smart ET methodology of the \blaze{} considers ETs as an intelligent wrapper
technology around a collection of highly optimized kernels that provide operation, data
type and architecture specific optimizations. A remarkable advantage of this methodology
is that this kernel based approach allows an easy integration of multi- and many-core
optimizations as well as GPU-based kernels.

\section{Inlining}
\label{sec:inlining}

Inlining is an essential issue for all ET-based frameworks: Without a complete inlining of the
entire ET functionality the expected performance level cannot be achieved. Therefore ETs are
vitally depending on the inlining capabilities of the used compiler. However, due to the
enormous number of nested function calls in ET codes the pressure on the compiler is very
high. Additionally, the \lstinline|inline| keyword is merely a recommendation for the compiler
to perform the inlining and not a binding instruction. Depending on the size of the function
the ETs are used in, the size of the compilation unit, the total number of instructions, etc.
the compiler might reject this recommendation and choose to insert function calls.

\begin{figure}[h!t]
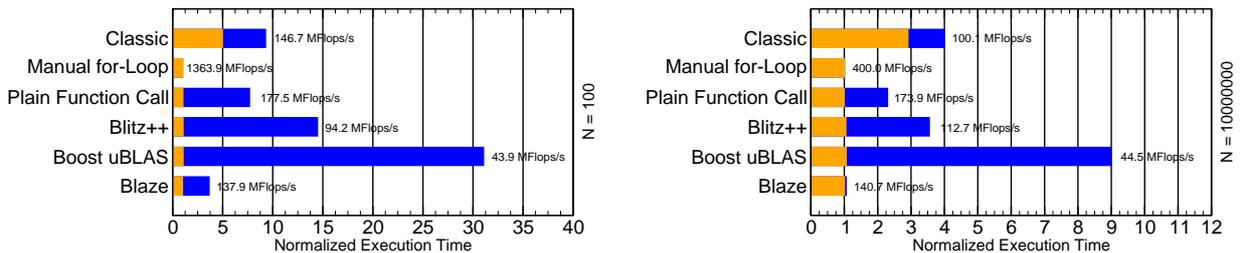

   \centering
   \includegraphics[clip=true,trim=0.5mm 0.9mm 0mm 0mm,angle=270,scale=0.29]{dvecdvecadd_100_noinline.eps}
   \hfill
   \includegraphics[clip=true,trim=0.5mm 0.9mm 0mm 0mm,angle=270,scale=0.29]{dvecdvecadd_10000000_noinline.eps}
   \caption[Performance comparison of the dense vector addition for proper and failed inlining]{
      \label{fig:dvecdvecadd_noinline}
      Performance comparison of the dense vector addition for proper and failed inlining.
   }
\end{figure}

During our performance measurements we frequently encountered problems with failed inlining, even
within apparently small test programs to measure the performance of a certain operation. Therefore
inlining seems to be a real issue that might result in bad performance although the implementation
would be able to deliver much more. In our measurements, we went to great lengths to ensure that
all ET functionality was properly inlined to measure the maximum possible performance. In order
to demonstrate the impact of failed inlining, however, Figure~\ref{fig:dvecdvecadd_noinline}
shows a comparison between proper and failed inlining in case of the dense vector addition
(the inlined performance values correspond to the results from Figure~\ref{fig:dvecdvecadd}).

%
%
%

As this comparison shows, inlining poses a severe and fundamental problem for all ET-based
production code\footnote{The authors have to admit that this also affects the ET implementation
of the \blaze{} library, but due to the concept of embedding HPC-kernels much less than the
other ET frameworks.}. Most importantly, programmers must not be overly confident in the
compiler's ability to (1) perform inlining to the required level and then (2) generate the
most efficient low-level loop code possible.



\section{Conclusion and Future Work}
\label{sec:conclusion}

There is very little ground for the reputation of standard Expression Templates to
be a performance optimization for array operations. They do achieve
their original goal of providing fast element-by-element array
arithmetic in combination with the benefits of high-level
constructs, because they effectively eliminate the generation of
temporaries in expressions. In this sense, they remedy a specific deficiency
of the C++ language. However, more complex operations
like BLAS level 2 and 3 procedures, sparse linear algebra, and generally
everything that profits from standard and architecture-specific low-level
optimizations, often show devastating performance levels. This is because
ETs are essentially an abstraction technique that hides
the details of actual data and operations types and reduces them to efficient
single-element access, which is insufficient:
We have shown that the widespread belief in advanced inlining and optimization
capabilities of C++ compilers is naive and unjustified. While aggressive
inlining is a necessary prerequisite for getting good performance
from ET source, it does not guarantee best low-level code.
There is no replacement for exploiting all possible knowledge about data
types, operations, and access patterns.

We have also introduced a new ET methodology, which we call ``Smart
Expression Templates.'' It eliminates the shortcomings of standard ETs
by reducing the ET mechanism to an intelligent wrapper around a
selection of highly optimized kernels or, in case of BLAS-type
operations, vendor-provided libraries. Smart ETs combine the advantages of a
domain-specific language (ease of use by high-level constructs,
readability, encapsulation, maintainability) with the performance
of HPC-suitable code. Moreover, they do not rely on aggressive
inlining as much as standard ETs do.

In this work we have restricted our discussion to sequential code.
Considering the importance of highly hierarchical,
multicore/multisocket building blocks in today's high performance
systems, a generalization of smart ETs to parallel computing on
distributed data structures seems natural and will be investigated.

\bibliographystyle{plain}
\bibliography{literature}

\end{document}